\documentclass[useAMS,usenatbib]{mn2e}

\usepackage{graphicx}
\usepackage{times}
\usepackage{natbib}
\usepackage{amsmath}

\renewcommand{\d}{\mathrm{d}}

\newcommand{\gtrsim}{\ga}
\newcommand{\lesssim}{\la}
\def\zsun{{\rm Z_\odot}}

\def\msun{{\rm M_\odot}}
\def\msunh{{\rm M_\odot/{\it h}}}

\def\Omegab{{\Omega_{0,\rm b}}}
\def\Omegadm{{\Omega_{0,\rm dm}}}
\def\Omegam{{\Omega_{0,\rm m}}}
\def\Omegal{{\Omega_{0,\rm \Lambda}}}

\title[SC rate]{Signatures of very massive stars: supercollapsars and their cosmological rate}
\author[U.~Maio \& M.V.Barkov]{
Umberto~Maio$^{1,2}$\thanks{E-mail: maio@oats.inaf.it,   \it Marie Curie Fellow}, Maxim~V.~Barkov$^{3,4,5}$\thanks{E-mail:
bmv@mpi-hd.mpg.de}\\
${}^1$INAF -- Osservatorio Astronomico di Trieste, via G. Tiepolo, 11, 34131 Trieste, Italy\\
${}^2$Leibniz-Institut f\"ur Astrophysik (AIP), An der Sternwarte 16, 14482 Potsdam, Germany\\
${}^3$Max-Planck-Institut f\"ur Kernphysik, Saupfercheckweg 1, 69117 Heidelberg, Germany\\
${}^4$Space Research Institute, 84/32 Profsoyuznaya Street, Moscow 117997, Russia\\
${}^5$Astrophysical Big Bang Laboratory, RIKEN, Saitama 351-0198, Japan
}

\begin{document}

\date{(draft)}
\pagerange{\pageref{firstpage}--\pageref{lastpage}}\pubyear{0}
\maketitle
\label{firstpage}

\begin{abstract}
We compute the rate of supercollapsars by using cosmological, N-body, hydro, chemistry simulations of structure formation, following detailed stellar evolution according to proper yields (for He, C, N, O, Si, S, Fe, Mg, Ca, Ne, etc.) and lifetimes for stars having different masses and metallicities, and for different stellar populations (population III and population II-I).
We find that supercollapsars are usually associated to dense, collapsing gas with little metal pollution and with abundances dominated by oxygen.
The resulting supercollapsar rate is about $10^{-2}\,\rm yr^{-1} sr^{-1}$ at redshift $z=0$, and their contribution to the total rate is $ < 0.1 $ per cent, which explains why they have never been detected so far.
Expected rates at redshift $z\simeq 6$ are of the order of $\sim 10^{-3}\,\rm yr^{-1} sr^{-1}$ and decrease further at higher $z$.
Because of the strong metal enrichment by massive, short-lived stars, only $\sim 1$ supercollapsar generation is possible in the same star forming region.
Given their sensitivity to the high-mass end of the primordial stellar mass function, they are suitable candidates to probe pristine population III star formation and stellar evolution at low metallicities.
\end{abstract}


%************************************************************************

\section{Introduction}\label{Sect:introduction}

%************************************************************************
%
%
Star formation is one of the main events taking place during the whole cosmic evolution.
At early times this process is particularly interesting because it happens in pristine regions where primordial stars can be formed via molecular-hydrogen cooling \cite[][]{SaslawZipoy1967} ruling gas collapse and fragmentation \cite[][]{Abel2002, Yoshida2003} 
independently from the background cosmological framework \cite[][]{Maio2006, Maio2011cqg}. Pollution of the surrounding environments with the first newly synthesised heavy elements boosts gas cooling capabilities through atomic fine-structure transitions \cite[e.g.][]{Maio2007} and leads from the primordial pristine `population III' (popIII) star formation regime to the subsequent metal-enriched `population II-I' (popII-I) one \cite[][]{Maio2010, Wise2012}. Nature and features of primordial stars are still matter of debate, since different works predict both very massive objects reaching  $\sim 10^2 - 10^3\,\rm M_\odot$ \cite[][]{Woosley2002, HegerWoosley2002, Heger2003}, and lighter ones close to standard values \cite[][]{Yoshida_et_al_2007, Greif2011, Stacy2013arXiv}.
Direct observational evidences of the first stars are not
available, yet, and their mass distribution is unknown, too. Despite these uncertainties, theoretical models of low-metallicity ($Z$) stars \cite[][]{Fryer2001} suggest that some of them could die explosively and leave a black hole as a remnant, possibly generating gamma-ray bursts (GRBs) \cite[][]{Bromm2006, Nuza2007, Kumar2008, Macpherson2013arXiv}.
Numerical simulations of cosmic star formation suggest that such explosions would affect directly the surrounding environments and would contaminate their host galaxy \cite[][]{Campisi2011, Maio2012, Salvaterra2013} with a number of different feedback effects \cite[][]{Maio2011b, PM2012, Wise2012, deSouza2013,BiffiMaio2013arXiv, deSouza2013arXiv}.
Moreover, GRBs could be optimal tools to study the status of the cosmic gas, its neutral, dust and metal content
\cite[e.g. discussion in][]{Salvaterra2013}, and to check the visible effects of different cosmological models\footnote{
  As e.g. non-Gaussianities \cite[][]{MaioIannuzzi2011, Maio2011, Zhao2013arXiv} and dark energy \cite[][]{Maio2006}.
}
in the primordial Universe \cite[][]{Maio2012}.
\\
Typical progenitors for GRBs have masses between $\sim 20-40\,\rm M_\odot$ up to $\sim 140\,\rm M_\odot$ ending their life as supernova (SN) explosions.
These dying stars can generate jetted $\gamma$ emissions with characteristic luminosities of $\sim 10^{50}-10^{51}\,\rm erg \, s^{-1}$ distributed from keV to MeV energy bands \cite[][]{Salvaterra2012}.
Stars with masses between $\sim 140\,\rm M_\odot$ up to $ 260\,\rm M_\odot$ could form pair instability supernovae (PISN)
\citep{Fraley68}, which are powerful explosions, but do not form jets. If the primordial stellar initial mass function (IMF) is top-heavy and samples very massive stars exceeding $\sim 260\,\rm \msun$, these latter could collapse into massive black holes and produce jets, giving birth to the commonly known supercollapsars (SCs).
As suggested by \cite{Komissarov2010}, the collapse of very massive stars could be accompanied by extremely energetic explosions with formation of magnetically driven jets \cite[e.g.][]{BZ1977,Barkov2008, Barkov2011}.
These jets can reach a power of about $10^{52}\,\rm erg/s$,
(i.e. at least ten times larger than the mentioned more common GRB progenitors)
releasing up to $\sim 10^{55}\,\rm ergs$ of the black-hole rotational energy \citep{Komissarov2010,MR10,SI11,TSM11} in about $10^4$~s.
Since primordial popIII stars are not expected to contribute to the GRB rate for more than $\sim 10$ per cent at $z\sim 6$ \cite[][]{Campisi2011, SYI11}, due to the strong metal spreading events after the first star formation episodes \cite[][]{Maio2010}, SCs should present an even lower contribution at these epochs.
\\
Unfortunately, the SC cosmological rate is still unknown and no SC occurrences have been measured by available experiments, so far. It is not clear whether this is due to the intrinsic nature of SCs or to the fact that stellar evolution models might be wrong in predicting properties or even existence of very massive stars -- i.e., SCs could not exist at all \citep{ydl12}. 
A typical feature which we could expect is that SCs should be located in recently collapsed gas, where the forming massive stars ($\sim 10^2\msun $) would explode more or less simultaneously with SC progenitors.
Since SCs are extremely sensitive to the high-mass end of the stellar mass distribution, their cosmological detection (or undetection) at different redshifts ($z$) could help place significant constraints on the IMF at various epochs and test the validity of stellar evolution models at low $Z$.
In order to do that, it is important to investigate the hosting metallicity environments and to estimate the rate of SCs and their contribution to the expected GRB rate in a cosmological context.
\\
In the next Sect. \ref{Sect:methods}, \ref{Sect:results} and \ref{Sect:conclusions} we will asses this problem via numerical, hydrodynamical, chemistry simulations following gas cooling at temperatures $\sim 10-10^{9}\,\rm K$, star formation, detailed stellar evolution according to proper stellar yields and lifetimes and supernova explosions from both popIII and popII-I regimes from high to low redshift. We will show that, indeed, SC events can be easily associated with first cosmic pollution events in the primordial Universe and that their rate can reach $\sim 10^{-2}\,\rm yr^{-1}\, sr^{-1}$ at lower redshift.
The consequent SC contribution to the total GRB rate results to be of $\sim 1$ per cent at $z\sim 6$ and drops down to $< 0.1$ per cent at $z\sim 0$.

%************************************************************************

\section{Methods}\label{Sect:methods}

%************************************************************************
%
%
\begin{figure*}
\centering
\includegraphics[width=0.28\textwidth]{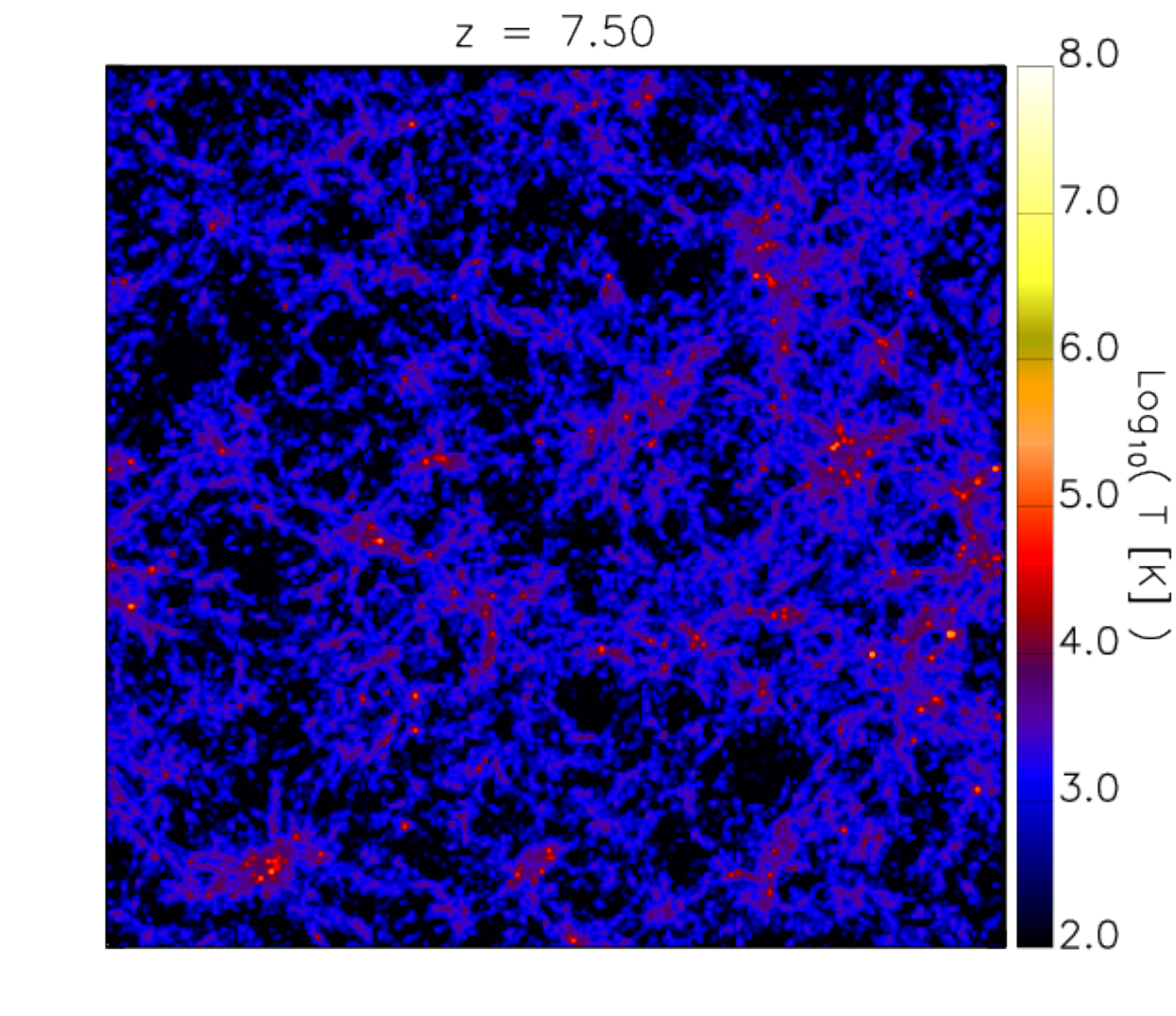}
\includegraphics[width=0.28\textwidth]{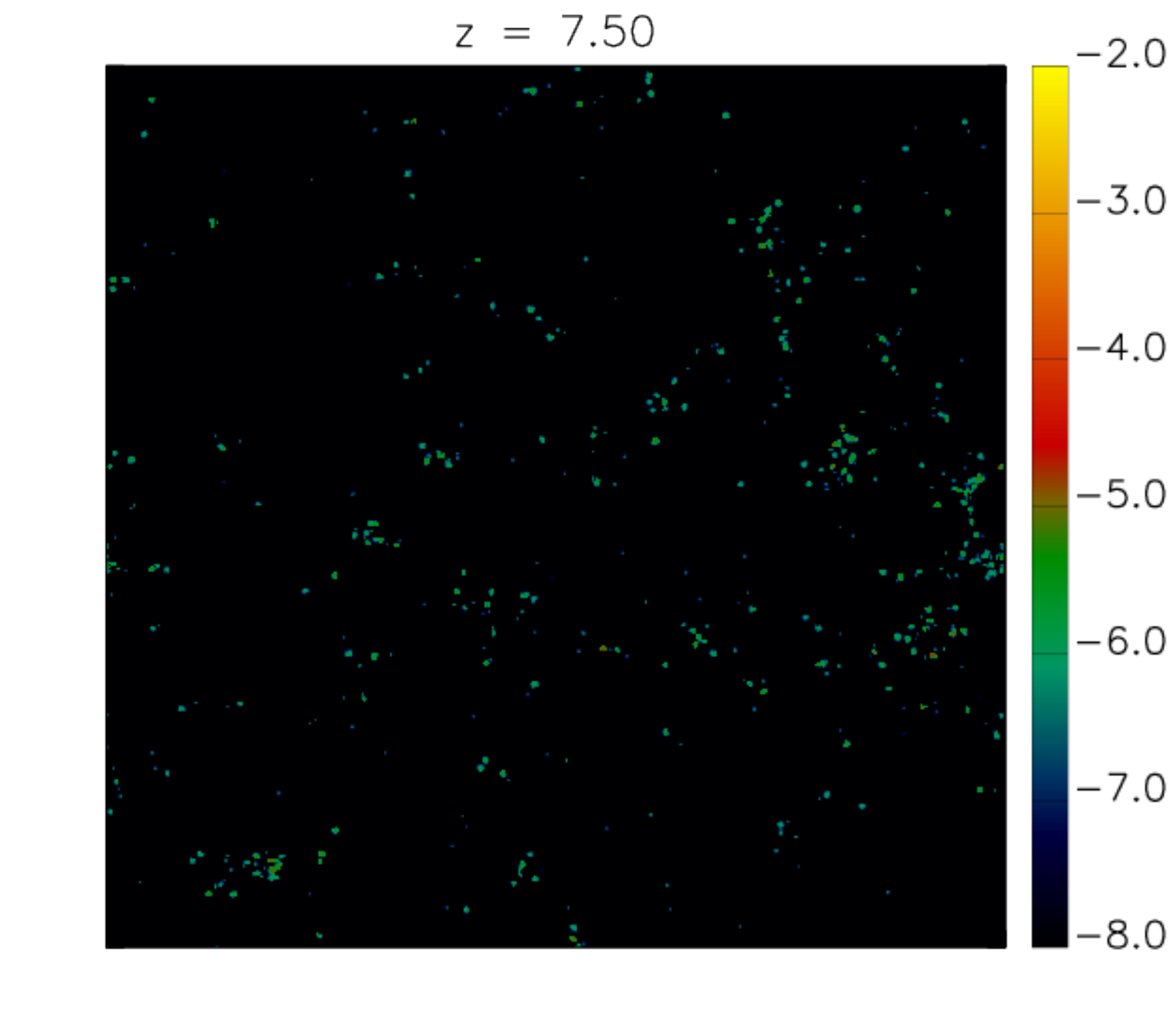}
\includegraphics[width=0.28\textwidth]{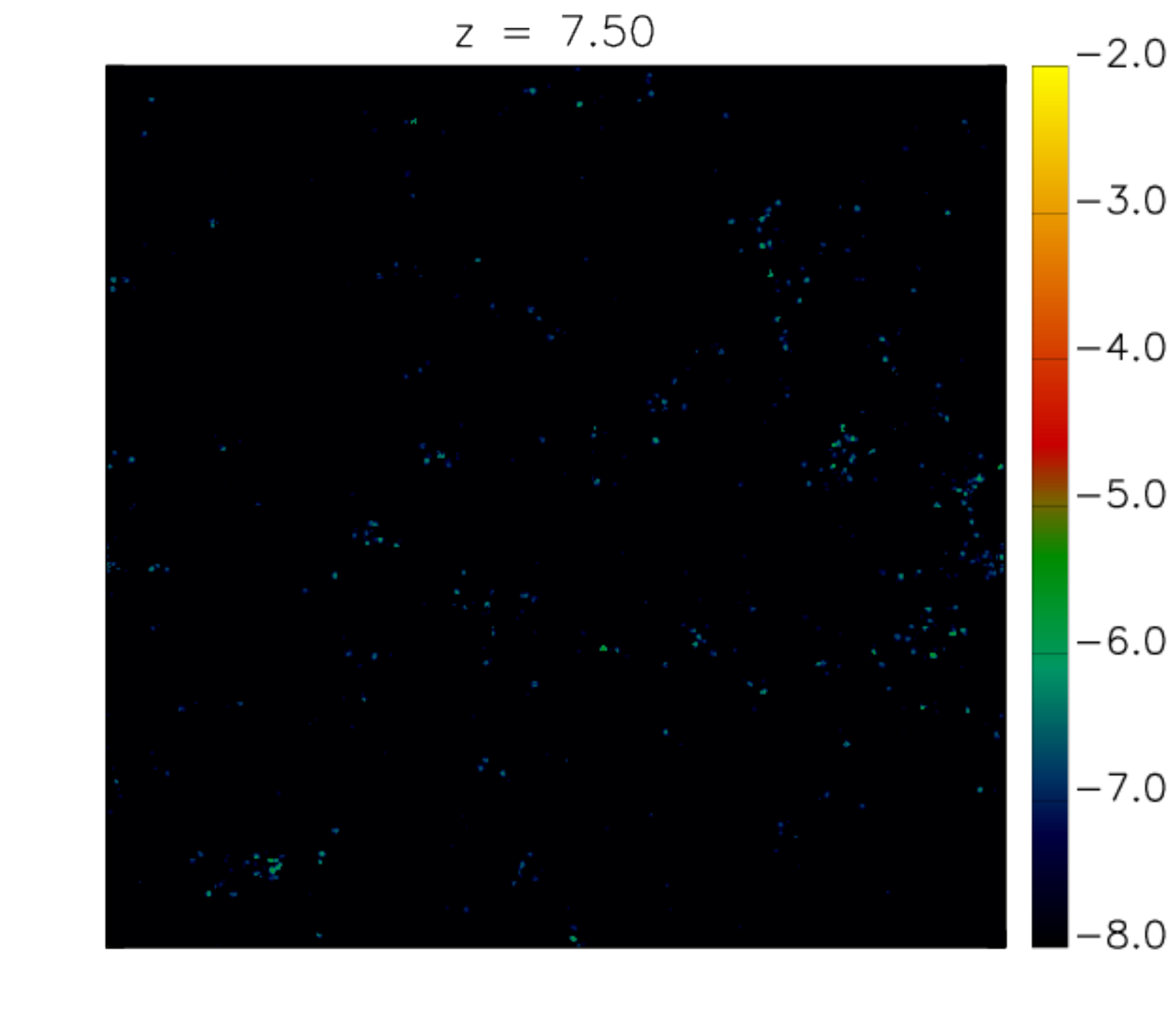}\\
\includegraphics[width=0.28\textwidth]{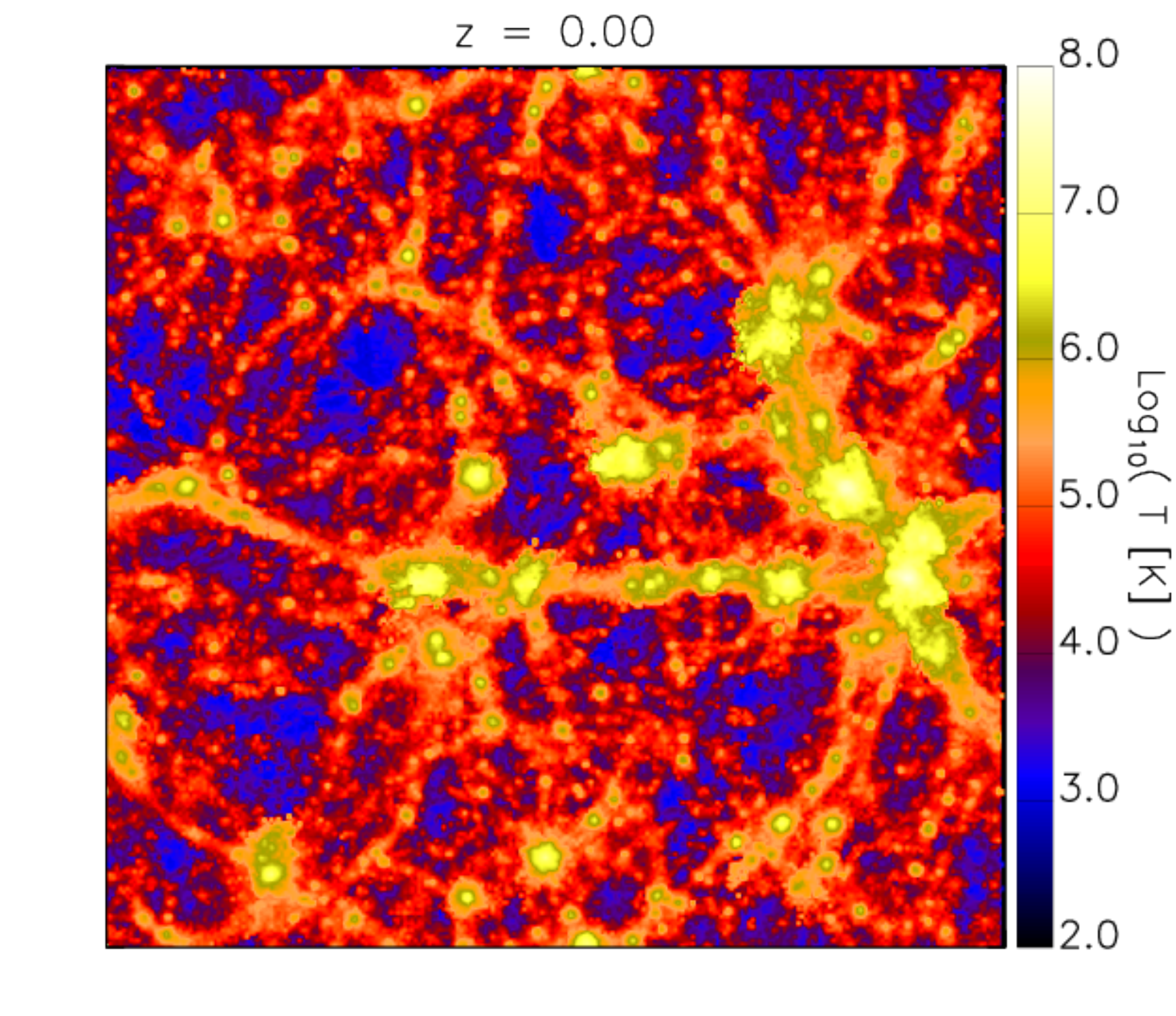}
\includegraphics[width=0.28\textwidth]{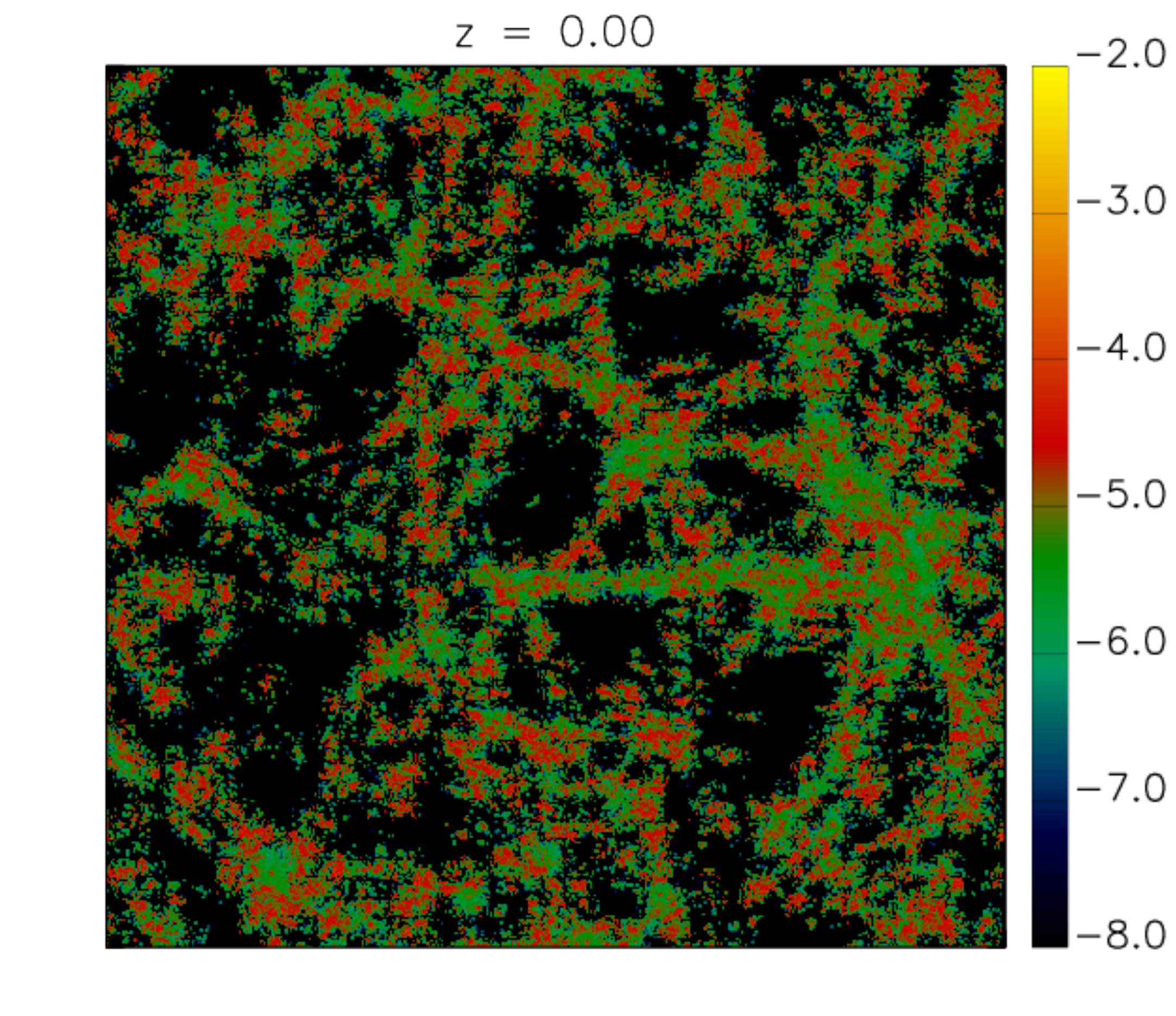}
\includegraphics[width=0.28\textwidth]{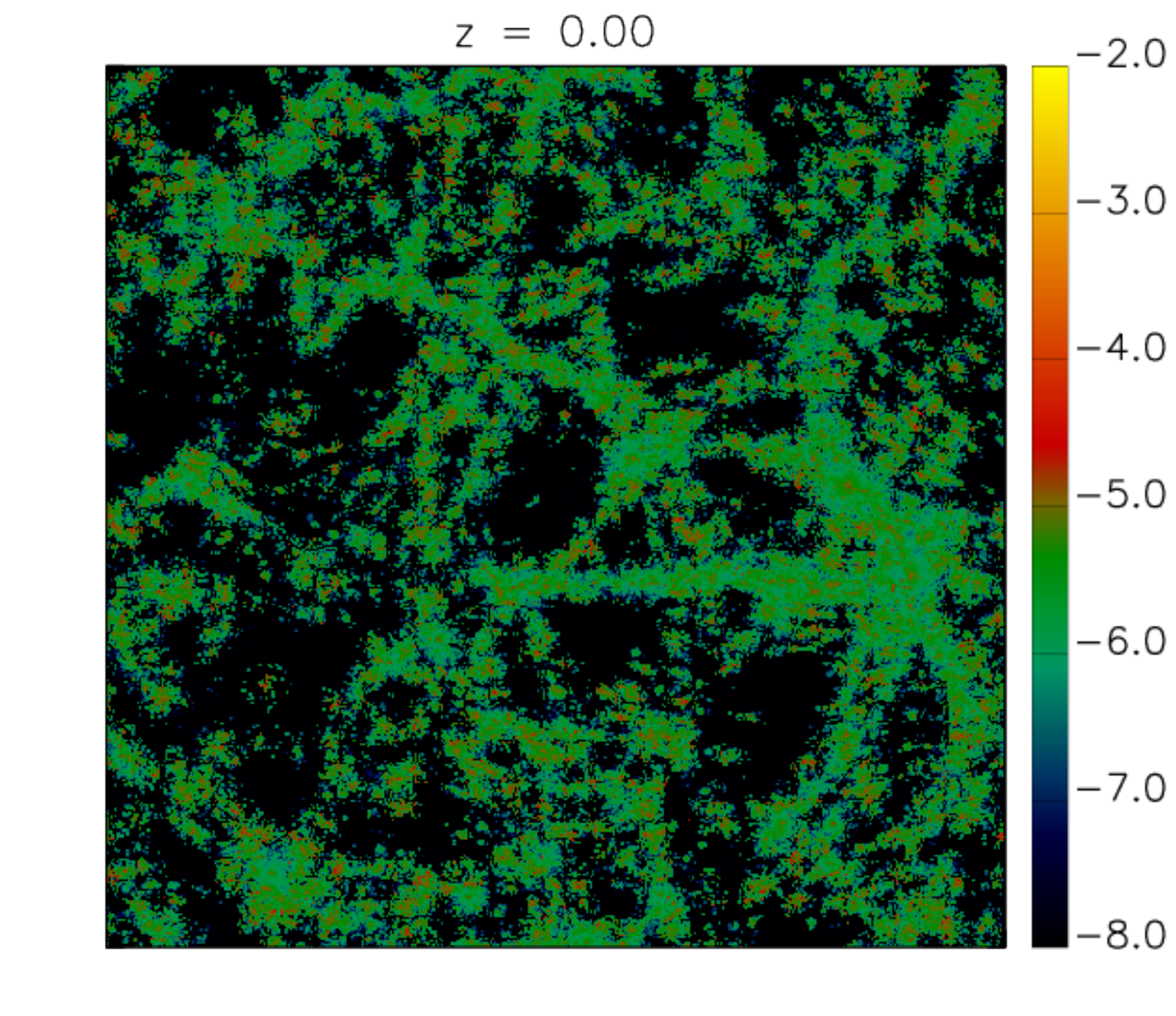}\\
{\bf
Temperature [K] \hspace{2.25cm} Oxygen (mass fraction) \hspace{2.25cm} Iron (mass fraction)\\
}
\caption[]{\small
  Maps of the gas temperature (left), Oxygen (center) and Iron (right) mass fractions at redshift $z=7.5$ (upper row) and $z=0$ (lower row).
At high-$z$ O-rich regions have just undergone massive-star explosions and, possibly, SC events.
At later times, enrichment patterns are degenerate with stellar evolution from all stellar masses.
}
\label{fig:maps}
\end{figure*}
We use numerical, hydrodynamical, chemistry simulations for structure formation in a flat $\Lambda$CDM Universe with cosmological-constant, total-matter, dark-matter and baryon-matter parameters
$\Omegal=0.7$,
$\Omegam=0.3$,
$\Omegadm=0.26$,
$\Omegab=0.04$,
respectively,
and with Gaussian initial matter perturbations in a box of 100~Mpc/{\it h} a side and corresponding gas mass resolution of $\sim 3\times 10^8\,M_\odot/h$
as described in \cite{MaioIannuzzi2011, Pace2013arXiv}.
They are performed with an updated version of Gadget-3 code \cite[][]{Springel2005} and implementing N-body and hydrodynamics calculations, coupled to gas cooling between $\sim 10$ and $\sim 10^9\,\rm K$ \cite[][]{Maio2007}, star formation and feedback mechanisms \cite[][]{Springel2003, Katz1996, Maio2009}, UV background \cite[][]{HaardtMadau1996}, and stellar evolution according to proper yields for He, C, N, O, Si, S, Fe, Mg, Ca, Ne, etc. \cite[][]{Tornatore2007, Maio2010} and delay times for stars with different masses and metallicities \cite[][]{Matteucci1986, Renzini1986, Schaerer2002}.
\\
The star formation rate density, $\dot\rho_\star$, resulting from the collapsing material, follows from gas density and thermodynamical 
state according to metal-dependent cooling and heating processes. Stars are formed according to an IMF, $\phi(M_\star)$, of the form
\begin{equation}
\phi(M_\star) \propto M_\star^x
\end{equation}
and defined over a suited stellar mass range.
More exactly, for the popII-I regimes we use a \cite{Salpeter1955}-like IMF over the range $\rm [0.1, 100] \msun$ with slope $x=-2.35$, 
while for the popIII regime a top-heavy IMF is assumed over the range $\rm [100, 500] \msun$ with slope $x=-2.35$. The transition from popIII to popII-I star formation takes place when the local $Z$ of the star forming region reaches a critical level \cite[][]{Bromm2003,Schneider2006, Santoro2006} of $Z_{crit}=10^{-4}\zsun$ \cite[uncertainties of this critical value in the range $10^{-6}\zsun-10^{-3}\zsun$
have negligeble impacts, as shown by][]{Maio2010}.
\\
Stellar evolution allows us to track the lifetimes of stars, $\tau_\star$, depending on the particular stellar masses, $M_\star$, normalized to the solar value, $m_\star \equiv M_\star / \msun$, and computed according to \cite[][]{Matteucci1986}:
\begin{eqnarray}
\tau_\star(m_\star) = \left( 1.2 m_\star^{-1.85} +  0.003 \right) {\rm Gyr}\\
{\rm for} \quad m_\star > 6.6 \nonumber
\end{eqnarray}
and  \cite[][]{Renzini1986}:
\begin{eqnarray}
\tau_\star(m_\star) = \left( 10^{ 1.338 - \sqrt{1.790 - 0.2232(7.764 - {\rm Log}\, m_\star)} } -9 \right) {\rm Gyr}\\
{\rm for} \quad m_\star \le 6.6. \nonumber
\end{eqnarray}
The lifetimes of primordial, massive, popIII stars are not very sensitive to $m_\star$ \cite[][]{Schaerer2002}:
\begin{equation}
\tau_\star \simeq 2\times 10^6 {\rm yr\quad for \quad} 100 \le m_\star \le 500.
\end{equation}
We stress that this basically means that we do not rely on the instantaneous recycling approximation (IRA), and heavy elements are released in the surrounding medium with delay times dictated by their parent stars.
Metal yields for popIII PISN, standard type II SN, AGB stars and type Ia SN are taken by 
\cite{HegerWoosley2002, Heger2010, WW1995, vdH1997,Thielemann2003}.
Feedback effects from exploding stars are responsible for enriching the medium with heavy elements and for heating the gas above $10^4\,\rm K$.
These processes are crucial for ionising H and for altering the cooling capabilities of the next generations.
\\
Once the star formation history is established \cite[see Fig. 4 by][]{MaioIannuzzi2011}, it is possible to compute the GRB rate and the 
SC rate at different redshfit. Following e.g. \cite{Campisi2011, Salvaterra2012, Salvaterra2013, Maio2012}, we assume that GRBs are good tracers of star formation\footnote{  This assumption is quite fair at low metallicities, where SCs are expected to form, however at $Z$ larger than $\sim 0.1\,\zsun$ there are still debates about possible biases.
}
and that the (comoving) GRB formation rate (GFR) number density, $\dot n_{\rm GRB}$, scales proportionally to the (comoving) star formation
rate density, $\dot \rho_\star$,
\begin{equation}
\dot n_{\rm GRB} = f_{\rm GRB} \, \zeta_{\rm BH} \, \dot\rho_\star,
\end{equation}
according to the fraction of BH formed per unit stellar mass, $\zeta_{\rm BH}$, and to the efficiency in powering a GRB during the 
collapse phase, $f_{\rm GRB}$. It is possible to compute the resulting (physical) rate, $R(z)$, as
\begin{equation}
\label{eq:rate}
R(z) = \frac{\d\dot N(z)}{\d\Omega}
\end{equation}
where $\d\Omega$ is the solid angle and the expected observable GRB number rate, $\dot N(z)$, is given by integrating $\dot n_{\rm GRB}$ over the cosmological volume and the luminosity function $\psi(L)$ above the sensitivity threshold, $L_{\rm th}$:
\begin{equation}
\label{eq:Ndot}
\dot N(z) = \gamma_b \int_z^{+\infty} \d z' \frac{\d V(z')}{\d z'} \frac{\dot n_{\rm GRB}(z')}{(1+z')} \int_{L_{\rm th}(z')} \psi(L') \d L' .
\end{equation}
Above we took into account cosmic time dilation by the factor $(1+z')^{-1}$ and the redshift dependence of the comoving cosmic volume element $\d V(z')$ and $L_{\rm th}(z')$.
The threshold sensitivity depends on the details of the particular kind of experiment considered (for the {\it Swift} instrument it corresponds to $ 0.4 \,\rm photons / s / cm^{2} $ in the [15, 150] keV energy band).
Morevoer, since GRBs are jetted sources, we also included the beaming factor, $\gamma_b$, to get a directly observable quantity
\footnote{
We warn the reader that sometimes alternative, but similar, definitions can be found in literature for $\dot N(z)$ and $R(z)$ and the number rate can be expressed e.g. per unit redshift interval or also per unit logarithmic redshift interval. The physical meaning remains almost unchanged, but the mathematical behaviour can demonstrate some variations.
}.
\\
The overall normalisation 
including all the pre-factors entering the equation above is consistently taken by the best-fit value in Table 1 by \cite{Maio2012} computed for a Gaussian initial matter distribution and a (comoving) box side of 100 Mpc/{\it h}.
It is estimated by using {\it Swift} data and by imposing that none of the GRBs observed by {\it Swift} were originated by popIII progenitors
\cite[as in e.g.][]{Campisi2011}.
Assuming that all the possible events could be detected by an ideally perfect instrument, independently from contingent limitations, equations (\ref{eq:rate}) and (\ref{eq:Ndot}) readily give us the estimated rate for both the entire population and for the supercollapsar one (having progenitors with masses $> 260 \rm M_\odot$).

%************************************************************************

\section{Results}\label{Sect:results}

%************************************************************************

\begin{figure*}
\centering
 \includegraphics[width=0.45\textwidth]{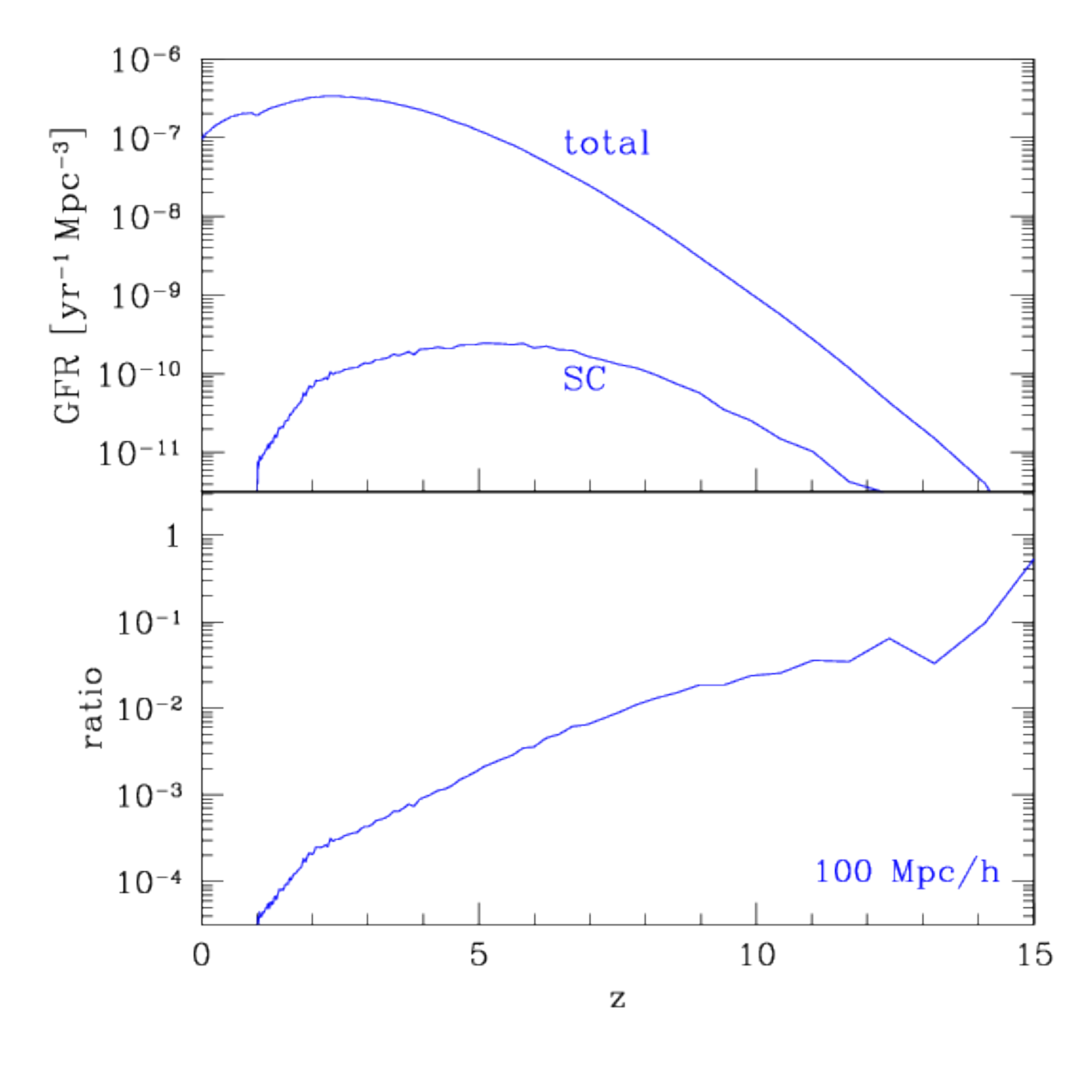}
\includegraphics[width=0.45\textwidth]{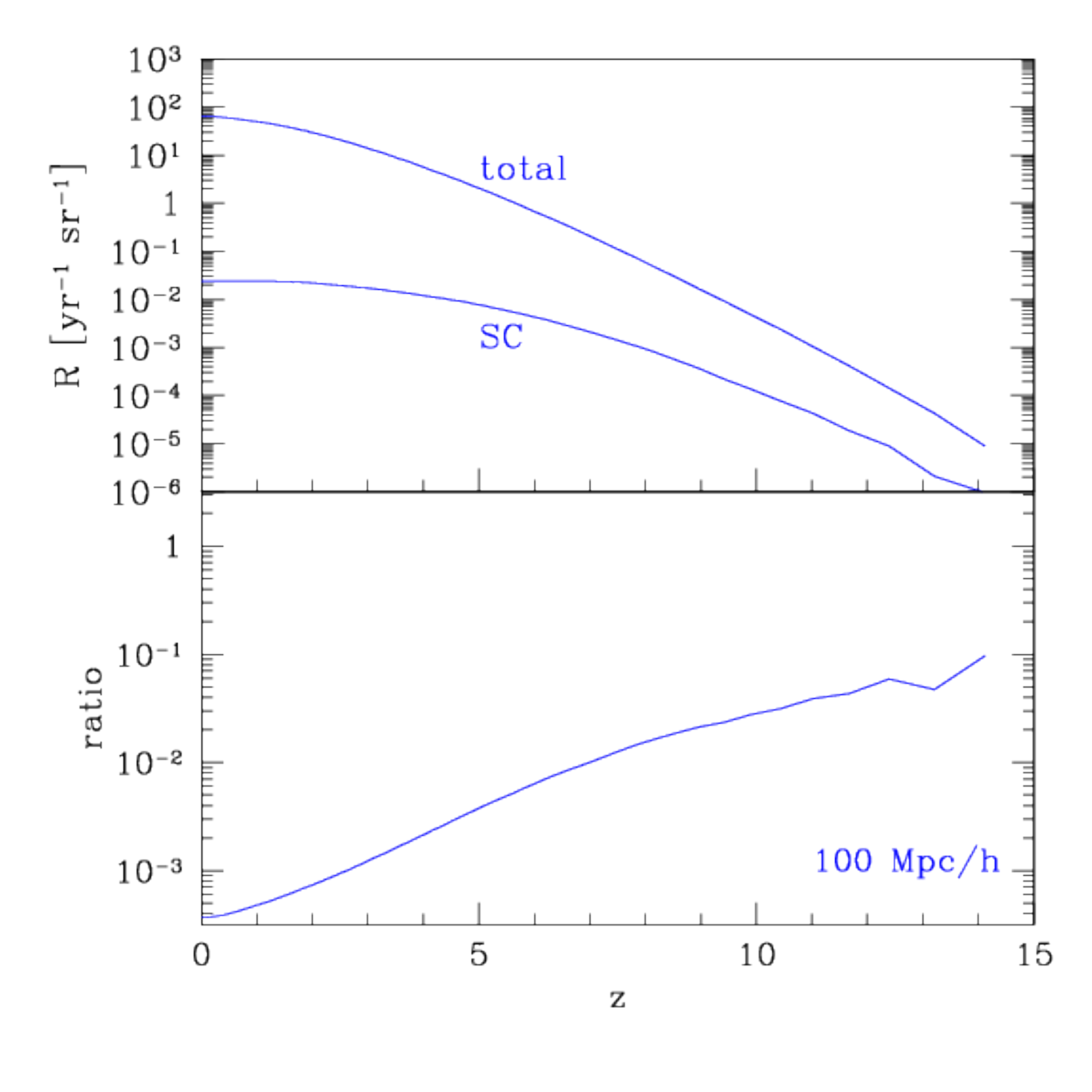}\\
\caption[]{\small
{\it Left}. Expected GRB formation rate number densities (top panel) according to the star formation rates of the whole population (total) and of the supercollapsar (SC) one, with corresponding SC fractional contribution (bottom panel) at different redshifts.
{\it Right}. Expected rates (top panel) for the whole gamma-ray burst population (total) and for supercollapsars (SC), with corresponding SC fractional contribution (bottom panel) at different redshifts.
}
\label{fig:rates}
\end{figure*}

In Fig.~\ref{fig:maps} we display some examples of theoretical expectations for gas temperature, $T$, and metal abundances both at high ($z=7.5$) and low ($z=0$) redshift. The overall temperature evolution is very clear, being $T$ mostly below $\sim 10^4\,\rm K$ in large volumes of the infant `dark' Universe. First star forming regions at $z >7.5$ are visible thanks to the higher temperatures and oxygen abundances locally determined by PISN and/or SN explosions and well recognisable at $z=7.5$ in haloes with typical sizes of $\lesssim 10^{11}\,\msunh$. At very high redshift, signals from $\lesssim 10^8\msunh$ structures might be hidden due to resolution limitations, and their effects on star and GRB formation would be visible at $z > 10$ \cite[as discussed in e.g.][]{Campisi2011}.
In the maps corresponding to $z=7.5$ metal pollution is still not very significant, because only a few massive stars could have exploded and spread O and $\alpha $ elements, leading to typical abundances of the order of $\sim 10^{-4}-10^{-2}\,Z_\odot$. Later on, the ongoing structure growth, star formation, metal pollution by stellar evolution, and feedback effects cause the Universe to be, in average, hotter and more metal rich, as shown by the $z=0$ maps, and consequently less suited to a massive-star regime and SC formation.
The production of the various heavy elements in the primordial Universe is still in its first stages, while at later epochs it gets more advanced and becomes particularly hard to find unpolluted, collapsing clumps.
Additionally, stellar evolution from various populations of stars with different masses and lifetimes plays a non-negligible role in affecting the yield elemental fractions, which is instead better recognisable at earlier epochs.
Late-time pollution is featured by increasing iron (Fe) production via low-mass, long-lived stars dying as type Ia SN.
In general, these results show a quite inhomogeneous enrichment, stronger in dense or clustered regions and weaker in peripheral areas of haloes or in voids. These environmental dependencies suggest that SCs are expected to be quite rare episodes, even if at higher $z$ conditions are definitely more favourable.
\\
We show our results for the total GFR and GRB rates and the corresponding SC sub-class in Fig.~\ref{fig:rates}.
On the left panel, we show the total GRB formation rate and the SC formation rate, as expected from the outcomes of our cosmological simulations.
In general, going from high to low $z$, the GFR increases, reaching a peak at $z\sim 2-3$ and following a subsequent decreasing trend down to redshift $\sim 0$. This behaviour is due to the fact that at the peak redshift the conditions are optimal for star formation over cosmological scales.
In fact, at higher $z$ structure growth is still in its initial stages and only rare high-$\sigma$ perturbations
could form and host gas cooling and stellar evolution processes. On the other extreme, low-$z$ cosmological evolution is dominated by an accelerated expansion (led by $\Omega_{\Lambda}$) which inhibits gas inflows and star formation mechanisms. A similar reasoning applies to SC formation rates, with the only difference that SCs are limited in unpolluted areas and, hence, their decline at low redshift is steeper.
Their peak is at $z\sim 5-6$ and features a quite broad evolution in time. The ratio between SC and total GFR is initially close to $\sim 1$, drops to $\sim 0.1$ at $z\sim 12-14$, and steeply decreases to $<10^{-2}$ at $z\sim 10$ and to $< 10^{-5}$ at $z\lesssim 1$.
On the right panel, we show the physical rate $R$ between $z=0$ and $z\simeq 15$. Since $R$ is an integrated quantity of a non-negative function in $z$, its trend is monotonic for both the total case and the SC case.
Due to the bursty nature of primordial galaxies \cite[][]{Salvaterra2013, BiffiMaio2013arXiv}, the total rate can be of the order of unity $\rm yr^{-1}\,sr^{-1}$ already at $z\gtrsim 6$, i.e. within the first Gyr of life of the Universe, while first star formation events could be traced back by GRBs and/or SCs up to much higher redshift ($z\sim 15$).
Given the paucity of $> 260\msun$ progenitors\footnote{
This is also related to the assumed IMF slope, however, by varying it within reasonable ranges one would get only slight changes around the same order of magnitude \cite[see previous discussion and parameter studies in][]{Maio2010}.
},
SCs are always sub-dominant with respect to the whole GRB population and their contribution rapidly decreases at lower redshift.
As previously mentioned, this is a direct consequence of the fact that, while cosmic structure formation and metal spreading proceed, the Universe becomes more and more enriched with metals and the dominant star formation mode is the popII-I one.
At earlier cosmological epochs the total GRB rate decreases with increasing $z$ because of the redshift suppression in equation~(\ref{eq:Ndot}) and the lower overall amount of stars formed. The resulting SC rate, which is only a few $10^{-2} \,\rm yr^{-1}\,sr^{-1}$ at $z\simeq 0$, makes SCs rare events.
For sake of comparison, we note that the total GRB rate at redshift $\sim 10$ turns out to be on the level of $\sim 3\times 10^{-3}\,\rm yr^{-1}\,sr^{-1}$, while at $z\sim 15$ it is only $10^{-5}\,\rm yr^{-1}\,sr^{-1}$. The corresponding SC rate results, instead, of the order of $10^{-4}$ and $10^{-6}\,\rm yr^{-1}\,sr^{-1}$, at redshift $z\sim $ 10 and 15, respectively. The ratio between SC and total rate is not constant, meaning that there is no direct proportionality between them, and that the non-linear effects of structure growth are affecting their behaviour, as well. In fact, the SC contribution decreases with redshift from $\sim 0.1$ at $z\sim 15$ down to $ < 10^{-3}$ at $z\sim 0$.
After a few times $10^{8}\,\rm yr$ ($z\sim 8-10$) from the onset of the first events, it drops to values around $\sim 10^{-2}-10^{-1}$, due to metal-enriched star formation and further standard popII-I GRB progenitors exploding. These conclusions are consistent with recent considerations relying on the fact that usually a large fraction  \cite[$\sim 93$ per cent;][]{BiffiMaio2013arXiv} of early structures gets metals via self-enrichment, independently by merger events or other external drivers, hence, only $\sim 1$ SC generation is usually possible in the same star forming region.
\\
An interesting detail of the SC formation rate is its relatively low-$z$ maximum position ($z\sim 5-6$), reflected by the corresponding change from a steeper to a shallower slope in $R$.
In this respect, semi-analytic investigations have tried to explore the popIII GRB epoch by relying on dark-matter mass functions and simple recipes for gas cooling, but their results have been shown to be quite sensitive to the underlying approximations with optimistic cases suggesting popIII GRB peak formation down to $z\sim 3$ \cite[][]{SYI11} and more extreme cases limiting popIII GRBs at $z > 5$ \cite[][]{Bromm2006}. It is intriguing to realize that our results about popIII GRBs, and more specifically about SCs, can be interpreted as a smooth, intermediate trend between these two opposite scenarios. This is possible thanks to the advantages of our numerical simulation that can simultaneously track, in a large volume, gaseous and chemical patterns of material in different haloes and with different thermodynamical properties, by following metal ejection and stellar death according to suited mass-dependent  lifetimes.\\
As a consequence, compared to previous assumptions \citep[e.g.][]{Komissarov2010, MR10}, where the SC peak was placed at $z\sim 10-20$, this implies that if SC will be observed, they should be bright transients in hard X-ray with duration of the order of several hours.
Unfortunately, it is very difficult to know the light-curve shape or spectra of SCs.
We can speculate that the closest analogue for a SC is a long GRB and that observational properties of SCs could be similar, but with a light curve stretched of a factor of 100 in time. The smoking gun for a SC event could be a pristine host galaxy with a relatively high column density of neutral gas and no or little pollution, possibly dominated by oxygen and $\alpha$ elements.
Such conditions would be met either in the primordial Universe or in pristine regions surviving cosmic metal enrichment at late times.
\\
We remind that the one outlined above represents the best-case scenario, derived by the undetection limits suggested by {\it Swift} and, hence, the actual values for the rates could be lower due to further unaccounted contingent or practical limitations while using real instruments.
Never the less, we can state that, if SCs existed, previous experiments could not have observed them because of insufficient statistical sampling.

%*****************************************************************************

\section{Conclusions}\label{Sect:conclusions}

%*****************************************************************************

We have computed the rate of supercollapsars (SCs) that could be originated by pristine star formation at different redshifts by means of cosmological simulations including N-body and hydro calculations, coupled to gas chemistry, cooling, star formation, feedback and stellar evolution according to proper metal yields and lifetimes \cite[see][]{Maio2010, MaioIannuzzi2011, Pace2013arXiv}.
\\
As a result of massive-star explosions, SCs can be easily associated with early star forming regions characterised by higher oxygen (O) and $\alpha$ elements pollution and lower iron (Fe) content, with O fractions reaching even 10 times Fe fractions. 
The supercollapsar rate is always a few orders of magnitude lower that that of the total GRB population and reaches values of a few $10^{-2}\,\rm yr^{-1}\, sr^{-1}$.
This is about 1/10 of what expected from numerical studies of popIII GRB populations \cite[][]{Campisi2011}.
The contribution of SCs to the total GRB rate regularly decreases in redshift from $\lesssim 1$ at high $z$ down to $ < 10^{-3}$ at $z\sim 0$, modulated by several physical mechanisms, as the ongoing cosmological structure growth, star formation processes, feedback effects, stellar evolution and metal spreading. Because of the strong metal enrichment by short-lived, massive stars, only $\sim 1$ SC generation is usually possible in the same star forming regions.
Despite their powerful jets and the large amounts of energy released, their extreme paucity justifies why previous searches did not find them.
Given their strong sensitivity to the high-mass end of the primordial IMF, they represent suitable candidates to probe pristine popIII star formation and stellar evolution at low metallicities. The detection of even one of such event can directly proof the existence of very massive popIII stars and pose constraints on their physical properties.

%*****************************************************************************

\section*{acknowledgments}
We acknowledge useful discussions with Ruben Salvaterra.
We are thankful to the Extreme Universe Laboratory and to the Institute for Nuclear Physics of the Moscow State University for kind hospitality during the finalisation of this work.
UM's research leading to these results has received funding from a Marie Curie fellowship by the European Union Seventh Framework Programme (FP7/2007-2013) under grant agreement n. 267251.
BMV acknowledges partial  support  by  RFBR  grant  12-02-01336-a.
We also acknowledge the NASA Astrophysics Data System and the JSTOR archives for their bibliographical support.

%*****************************************************************************

\bibliographystyle{mn2e}
\bibliography{bibl.bib}

\label{lastpage}
\end{document}